\documentstyle[psfig]{aipproc}

\def\ApJ#1{{\it Ap. J.} {\bf #1}}
\def\ApJLett#1{{\it Ap. J. Lett.} {\bf #1}}
\def\MNRAS#1{{\it M.N.R.A.S} {\bf #1}}
\def\Nature#1{{\it Nature} {\bf #1}}

\begin{document}
\title{Properties of GRB Host Galaxies}
\author{Dieter H. Hartmann$^1$ and David L. Band$^2$}
\address{$^1$Department of Physics and Astronomy,
Clemson University, Clemson, SC 29634\\
$^2$CASS, UC San Diego, La Jolla, CA 92093}
\maketitle
\begin{abstract}
The transients following GRB970228 and GRB970508 showed that these
(and probably all) GRBs are cosmological.  However, the host galaxies
expected to be associated with these and other bursts are largely
absent, indicating that either bursts are further than expected or the
host galaxies are underluminous.  This apparent discrepancy does not
invalidate the cosmological hypothesis, but instead host galaxy
observations can test more sophisticated models. 
\end{abstract}
\section*{The Absence of the Expected Host Galaxies}
Observations of the optical transients (OTs) from GRB970228
\cite{jan97} and GRB970508 \cite{bond97} have finally provided the
smoking gun that bursts are cosmological.  In most cosmological models
bursts occur in host galaxies:  are these galaxies present, and
conversely, what can we learn from them?  Underlying any confrontation
of theory and data must be a well defined model.  Here we show that 
the host galaxy observations are not consistent with the expectations 
of the simplest cosmological model, and that these observations can be 
used to test more sophisticated models.

In the simplest (``minimal'') cosmological model the distance
scale is derived from the intensity distribution logN--logP assuming
bursts are standard candles which do not evolve in rate or intensity.
Bursts occur in normal galaxies at a rate proportional to a galaxy's
luminosity.  This model predicts the host galaxy distribution
for a given burst. Are the expected host galaxies present? 

For GRB970228 an underlying extended object was found \cite{jan97},
but its redshift and nature have not been established. If the observed
``fuzz'' is indeed a galaxy at $z\sim\onequarter$, it is $\sim$5
magnitudes fainter than expected for a galaxy at this redshift. For
GRB970508 no obvious underlying galaxy was observed~\cite{pian97} and
the nearest extended objects have separations of several arcseconds,
but spectroscopy with the Keck telescopes \cite{metzger97} led to the
discovery of absorption and emission lines giving a GRB redshift of $z
\ge 0.835$. The {\it HST} magnitude limit $R_{\rm lim}\sim 25.5$
\cite{pian97} for a galaxy coincident with the transient again
suggests a host galaxy fainter than expected. Similar conclusions
follow from the inspection of IPN error boxes
\cite{BH98,schaefer92,schaefer97}, but see also
\cite{larson97,vrba97}. This absence of sufficiently bright host
galaxies is often called the ``no-host'' problem, which is a misnomer.
The point simply is that if galaxies such as the Milky Way provide the
hosts to most bursters, and if their redshifts are less than unity, as
predicted by the minimal model, we expect to find bright galaxies
inside a large fraction of the smallest IPN error boxes. 

To demonstrate this quantitatively, consider the apparent magnitude of
a typical host galaxy, which we assume has $M_*(B) = -20$
(approximately the absolute magnitude of an $L_*$ galaxy---see
discussion below). Using Peebles' notation \cite{peebles}, the
apparent magnitude is 
\begin{equation}
m = 42.38 + M + 5\ {\rm log}\left[y(z)(1+z)\right] + K(z) + E(z) + 
   A(\Omega,z) + \chi(z) \ \ ,
\end{equation}
where $K(z)$ is the usual K-correction, $E(z)$ corrects for the
possible evolution of the host galaxy's spectrum, $A$ is the sum of
Galactic foreground (position dependent) and intergalactic extinction,
and $\chi(z)$ represents any corrections that apply in hierarchical
galaxy formation scenarios, where galaxies are assembled through the
merger of star forming subunits. The commonly found term $5\log(h)$ is
already absorbed in eq.~(1). Neglecting potentially large corrections
from the $K$, $E$, $A$, and $\chi(z)$ terms, a host like the Milky Way
with $M\sim -20$ would have an apparent magnitude $m\sim 22$ for
redshifts of order unity. Several small IPN error boxes have no galaxy
of this magnitude or brighter. Our simplified treatment agrees with
Schaefer's conclusion\cite{schaefer92,schaefer97} that typical
galaxies at the calculated burst distance are absent from burst error
boxes. 

Thus bursts are further than predicted from the logN--logP
distribution without evolution, or they occur in underluminous
galaxies; an extreme limit of the latter alternative is that bursts do
not occur in galaxies. 
\section*{Host Galaxies as a Probe of Cosmological Models}
The search for host galaxies is a powerful test of cosmological burst
models.  From the two above mentioned OTs we conclude that GRBs are
cosmological, but the observations have not fixed the distance scale
quantitatively, nor have they determined the energy source. While the
x-ray, optical, and radio lightcurves (for GRB970508 only) are
consistent with the predictions of the basic ``fireball afterglow''
picture, the fireball's central engine could be the merger of a
neutron star binary, the collapse of a massive, rotating star, or the
jet produced by accretion onto a massive black hole residing at the
center of an otherwise normal galaxy.  

The host galaxies found within burst error boxes are a powerful
discriminant between different models for the burst energy source.
Note that the region within which a host would be acceptable
surrounding the sub-arcsecond localizations of an OT is effectively
the error box for the host galaxy. Almost all models assume that
bursts are associated with galaxies; the issue is the relationship
between the burst and the host.  In models such as the momentary
activation of a dormant massive black hole the burst rate per galaxy
is constant. On the other hand bursts are an endpoint of stellar
evolution in most models, and therefore to first order we expect the
burst rate per galaxy in these models to be proportional to the
galaxy's mass and thus luminosity.  These two model classes have
different host galaxy luminosity functions $\psi(M)$ with different
average values of $M$ (the absolute magnitude).  In the first case,
$\psi(M)$ is proportional to the normal galaxy luminosity function,
while in the second case $\psi(M)$ is proportional to the normal
galaxy luminosity function weighted by the luminosity $L\propto
10^{-0.4M}$.  We approximate the normal galaxy luminosity function
with the Schechter function: 
\begin{equation}
\Phi(M) = \kappa\ 10^{0.4(M_*-M)(\alpha+1)}\ \left[{\rm exp}\left(-10^{0.4
(M_*-M)}\right)\right]
\end{equation}
where $\kappa$ is the normalization, $\alpha$ is the slope of
the faint end, and $M_*$ is the absolute magnitude of an $L_*$ galaxy.
Here we use $\alpha=-1$. In the B band $M_*(B)= -19.53$, which
corresponds to $L_*(B) = 1.8\times 10^{10} L_\odot \ h_{75}^{-2} \sim
3\times 10^{11}\ L_{\odot}(B)\ h_{75}^{-2}$. In Figure~1 
\begin{figure}[b!] 
\centerline{\psfig{file=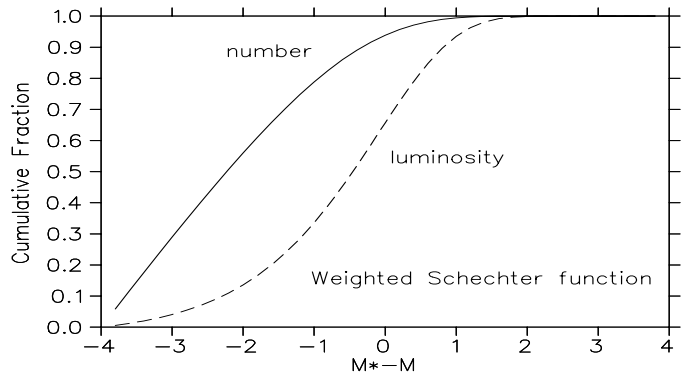,height=1.9in,width=3.5in,angle=90}}
\vspace{10pt}
\caption{The cumulative distribution of host galaxy magnitudes if
their luminosity function is weighted by number or luminosity. If GRBs
trace light, the 50$\%$ point is near $M_*$. If it is proportional to
galaxy number a typical host would be $\sim2$ magnitudes fainter.} 
\label{fig1}
\end{figure}
we show the cumulative distributions for the host galaxy magnitudes
for the two model classes. As can be seen, the average host galaxy
magnitude (i.e., at 0.5) differs by $\sim1.75$ magnitudes. 

However, we can make better predictions about the host galaxies in
cosmological models where bursts are a stellar endpoint.  In such
models, the burst rate should be a function of the star formation rate
(SFR).  If there is a substantial delay (e.g., of order a billion
years or more) between the GRB event and the star forming activity
that created the progenitor, then the burst rate integrates over a
galaxy's SFR, and we would not expect the host galaxy to display the
signatures of recent star formation.  Furthermore, if the progenitor
is given a large velocity, then it may travel a large distance from
the host galaxy before bursting, and it may become impossible to
associate a galaxy with the burst. 

In many models the burst occurs shortly after its progenitor star
forms (e.g., within a hundred million years or less).  We would then
expect that on average bursts would occur in galaxies showing evidence
of recent star formation.  The burst rate should be proportional to
the SFR, both for individual galaxies and for a given cosmological
epoch. 

In particular, the burst rate and the SFR should have the same
history, as was recently considered by several groups
\cite{vahe97,sahu97,totani97,wijers97}. Extensive redshift surveys and
data from the Hubble Deep Field have reliably determined the cosmic
star formation history to $z\sim5$
\cite{connolly97,ellis97,lilly97,madau96,madau97}. The data clearly
suggest a rapid increase in the comoving SFR density with increasing
redshift, SFR $\propto (1+z)^4$, reaching a peak rate (at $z\sim1.5$)
about 10--20 times higher than the present-day rate, and decreasing
slowly to the present value by $z\sim5$. This evolution function,
$\eta(z)$, enters the differential rate vs. (bolometric) peak flux 
\begin{equation}
\partial_P{R} \propto P^{-5/2}\ {\rm E}(z)^{-1}\ \eta(z)\ (1+z)^{-3}\
\left[(1+z)\partial_zy(z) + y(z)\right]^{-1} \ \ ,
\end{equation}
where E($z$) and $y(z)$ are defined in \cite{peebles}. For small
redshifts the logarithmic slope of this function is Euclidean, i.e.
$-$5/2.  The solid curve of Figure~2 
\begin{figure}[b!] 
\centerline{\psfig{file=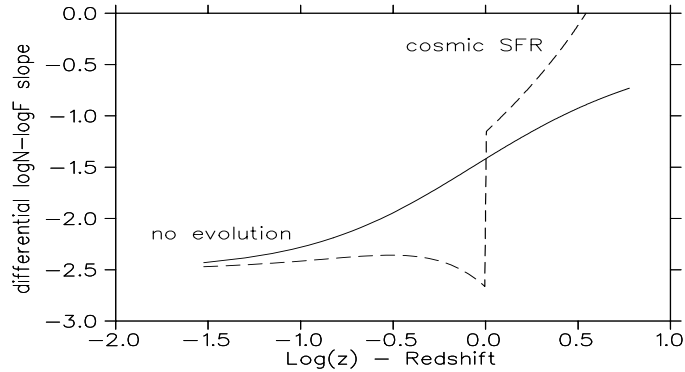,height=1.9in,width=3.5in,angle=90}}
\vspace{10pt}
\caption{The slope of the differential GRB brightness distribution.
Non-evolving sources (solid curve) quickly show significant deviation
from the Euclidean value $-$5/2 with increasing redshift. If the GRB
rate is proportional to the SFR (see text) the apparent Euclidean
slope extends to $z\sim 1$ (dashed curve). For greater $z$ the
geometry of the universe together with a now decreasing burst rate
cause the slope to deviate rapidly from $-$5/2.} 
\label{fig2}
\end{figure}
shows the effects of geometry (bending of logN--logP) and the dashed
curve demonstrates how $\eta(z)$ compensates for the geometry out to
the redshift at which the cosmic SFR peaks. At larger redshifts the
effects of geometry and decreasing SFR then combine and the slope
flattens quickly. Comparison with BATSE data suggests that this SFR
model deviates from the pseudo-Euclidean slope too abruptly. While
several studies \cite{sahu97,totani97,wijers97} report that the
observed SFR generates a brightness distribution consistent with BATSE
data, our findings support the different result of Petrosian $\&$
Lloyd \cite{vahe97}, who suggest that other evolutionary effects must
be present in addition to the density evolution described by
$\eta(z)$. While a good fit to the data requires a more sophisticated
model of source evolution, the basic message is likely to be the same:
the logN--logP distribution does not exclude GRB redshifts much
greater than unity. 

Therefore, the absence of the host galaxies predicted by the
``minimal'' cosmological model does not call the cosmological origin
of bursts into question.  Instead, host galaxy observations will teach
us where bursts occur.

\end{document}